\documentclass{article}

\usepackage{graphicx}
\usepackage[round]{natbib}
\usepackage{epsfig}

\title{
A new method for making objective probabilistic climate forecasts from numerical climate models
based on Jeffreys' Prior
}

\author{Stephen Jewson\footnote{\emph{Correspondence email}: \texttt{stephen.jewson@rms.com}},
Dan Rowlands and Myles Allen\\}

\begin{document}
\maketitle

\begin{abstract}
We argue that it would be desirable to use Jeffreys' priors in the construction
of numerical model based probabilistic climate forecasts, in order that those forecasts could be
argued to be objective. Hitherto, this has been considered computationally unfeasible.
We propose an approximation that we believe makes it feasible, and derive closed-form expressions
for various simple cases.
\end{abstract}

\section{Introduction}

There are many reasons for trying to predict future climate, ranging from
the assessment of insurance risk to the determination of appropriate government policy.
These different uses of climate forecasts need predictions of different aspects of climate, on different time-scales, and
this has led to the use of a range of different methods for making climate predictions.
For instance, the assessment of insurance risk is typically based on predictions of extreme climate over lead times of
just a few years, and such predictions are generally made using an appropriate blend of statistical
and numerical methods.
The planning of government policy, on the other hand, is typically based on predictions of mean climate over lead times of
many decades, and such predictions are generally made using numerical climate models, such as those described in the
IPCC fourth assessment report~\citep{IPCC4}.

In both statistical and numerical climate forecasting, attempts are being made to understand and incorporate all sources of uncertainty into forecasts.
If we split uncertainty into aleatoric (irreducible) and epistemic (potentially reducible) components,
then estimating the aleatoric uncertainty (often known as randomness or variability in statistical
modelling, and initial condition uncertainty in numerical climate modelling) is typically the easier of the two.
In statistical model climate predictions aleatoric uncertainty can be estimated directly from data,
and in numerical model climate predictions it can be estimated by creating initial condition ensembles.
Estimating epistemic uncertainty (often separated into model and parameter uncertainty) is typically more difficult.
In statistical modelling Bayesian methods can be used to estimate parameter uncertainty, and various different methods
have been proposed for incorporating model uncertainty (although none are particularly satisfactory).
In numerical modelling a number of approaches for estimating parameter uncertainty have been proposed and tested, such as those described in~\citet{frame05}, ~\citet{tomassini07} and~\citet{allen09}, and model uncertainty
has been estimated by using multimodel ensembles, such as that described in~\citet{cmip}.

In this article we consider the estimation of the parameter uncertainty component of the epistemic
uncertainty in numerical climate models. Methods proposed at this point have various advantages
and disadvantages.
The classical statistical methods used in~\citet{allen09} have the advantage that they
avoid the use of subjective priors, but they do not give a method for producing a probabilistic forecast,
and uncertainty is only represented by confidence intervals around best guess predictions.
The subjective Bayesian methods described in~\citet{frame05} and~\citet{tomassini07} do produce probabilistic forecasts, but the use of subjective priors introduces arbitrariness into the forecast.
This arbitrariness is fine if the users of the forecast are prepared to accept the priors that have been used.
On the other hand, it would be perfectly reasonable \emph{not} to accept the proposed priors, and the resulting forecast, since they are essentially arbitrary rather than scientifically determined.
Such forecasts can never, therefore, be expected to lead to much of a consensus about likely future climate states, unless the data starts to overwhelm the prior (which appears not to be the case for now).
For policy making this is unfortunate, since lack of consensus on the underlying forecasts is likely to lead to lack of
consensus on the appropriate course of action.
Another disadvantage of subjective priors is that forecasts made using such priors can never be backtested in an honest
way, since it would never be possible to argue that the prior had not been formulated using data from the testing
period.
With climate forecasts based on subjective priors it will never be possible, therefore, to make the standard mathematical modelling argument that one should believe
forecasts of the future because forecasts of the past performed well. Instead, the belief (or lack of belief) that forecasts of the future are likely to be accurate has to be based solely on faith in the modelling process (or lack of faith), rather
than a combination of faith in the modelling process and empirical demonstration of predictive ability.

To make up for this lack of methods for producing non-arbitrary probabilistic
climate forecasts we are embarking on an attempt to understand how to use Jeffreys' Priors to produce
probabilistic climate forecasts.
Jeffreys' Priors are the standard conventional prior used in situations where it is preferable to avoid
including subjective information.
They thus offer the hope of being able to achieve a greater level of consensus among scientists as to the distribution of future climate states. In addition, they allow for the possibility that climate models could be back-tested in
an honest way, which, if such back-testing indicates that climate models can make good out-of-sample forecasts,
should lead to greater confidence in climate model predictions of the future.

Jeffreys' Priors are already being used in statistical climate forecasts used in industry~\citep{j115}.
One of the reasons that they have not been used to date in numerical model forecasts is that they have been considered to be too computationally demanding (see the comments in~\citet{tomassini07}, page 1243).
We are hoping to be able to prove that this is not the case, through the use of judicious approximations,
and through the use of distributed computing available as part of the climateprediction.net project.

In section~\ref{s2} below we introduce Jeffreys' Priors.
In section~\ref{s3} we then discuss an approximation that can be made when calculating Jeffreys' Priors for numerical climate models, that we believe renders them practical.
Finally in section~\ref{s5} we summarise and discuss our findings.

\section{Jeffreys' Priors}\label{s2}

We will work within a notational framework in which we have historical climate data $x$, and we are trying to make a probabilistic prediction of
future climate data $y$. We write this prediction as $p(y|x)$. Given a model for this probability distribution (which could, at this point,
be a statistical model or a numerical model ensemble), we can make a prediction using:
\begin{equation}
    p(y|x)=\int p(y|\theta) p(\theta|x) d \theta
\end{equation}
where $\theta$ is the parameter vector of the model.

This equation says that our prediction is going to be made up of a weighted average of predictions $p(y|\theta)$ from models with different parameter values,
weighted by the probability of each value of the parameter given the data, $p(\theta|x)$.
Using Bayes theorem, we can factorise $p(\theta|x)$, giving:
\begin{equation}\label{eq2}
    p(y|x) \propto \int p(y|\theta) p(x|\theta) p(\theta) d \theta
\end{equation}
$p(x|\theta)$ is known as the likelihood (in both classical and Bayesian statistics) and can be evaluated by comparing the performance of the
various models with data. $p(\theta)$ is the prior.
~\citet{tomassini07} and others have used subjective priors, based on expert judgement.
To minimise the arbitrariness, or subjectiveness, of our forecasts we, however, would like to choose the prior in a non-arbitrary way.
Setting the prior to a constant is not an option, since a constant in one coordinate system may not be a constant in another coordinate system.
The only practical solution currently available is Jeffreys' Prior, defined as constant for parameters that represent a shift,
$1/\sigma$ for parameters $\sigma$ that represent a scaling, and for other more general parameters $\theta$
(where $\theta$ can be either a single parameter, or a parameter vector) as:
\begin{equation}
    p(\theta)=\sqrt{-\mbox{det}\left[\mbox{E}\left(\frac{\partial^2 \ln p}{\partial \theta_i \partial \theta_j}\right)\right]}\\
\end{equation}

where $p=p(x|\theta)$, and the expectation is over all possible values of $x$.
We note as a warning to readers who may not have come across Jeffreys' Prior before that this expression
is somewhat difficult to understand.
In particular we note that the quantity $p(x|\theta)$ is the likelihood function of the model
\emph{for arbitrary $x$}, as distinct from the $p(x|\theta)$ that occurs in equation~\ref{eq2} above,
which is the same likelihood function, but with $x$ set to the observed values.
The expectation operator $E$ is an integral over all possible values for $x$ that could have
occurred in the past (i.e. is a typical classical statistical expectation). It does not
commute with the derivative of $\ln p$.

Jeffreys' prior was originally presented in~\citet{jeffreys}, and has been widely used since.
The main attraction of Jeffreys' Prior is that
it has the property that it is invariant under coordinate
transformation of $\theta$: the final prediction does not depend on the coordinates $\theta$ that are chosen to parametrise the model. The proof of this is standard, but typically only given in very abbreviated form.
For completeness, and clarity, we include the proof in the appendix, both in the single parameter form
(in appendix 1) and in the multiple parameter form (in appendix 2).
Jeffreys' Prior also has many other interesting properties, that have been widely discussed in the
statistics literature (see, for example, ~\citet{bernardo}).


\section{Approximations to Jeffreys' Priors for use in climate modelling}\label{s3}

How, then, might we evaluate Jeffreys' prior for a climate model?
First, we note that Jeffreys' prior is only a function of the model, and not of the observational data. So evaluating
Jeffreys' prior is `simply' going to be a question of running the climate model a number of times, in the right way, and processing the output.
Given careful experimental design, the integrations needed to calculate Jeffreys' Prior could be the same
as those needed to calculate the likelihood term $p(x|\theta)$ in equation~\ref{eq2}, thus minimizing
the computational effort required.
The obvious brute-force approach to evaluating Jeffreys' prior would then be:

\begin{itemize}
  \item Run initial condition ensembles on a parameter grid to estimate $p(x|\theta)$
  (with one initial condition ensemble for each value of $\theta$).
  \item Numerically differentiate $p(x|\theta)$ to give $\frac{\partial^2 \ln p}{\partial \theta^2}$
  \item Numerically take the expectation, to give $E\left[\frac{\partial^2 \ln p}{\partial \theta^2}\right]$
  \item Take the square root, at each value of $\theta$.
\end{itemize}

The use of emulators (aka response surfaces) can probably help in the estimation of
$p(x|\theta)$, but nevertheless this approach is likely to be computationally challenging,
given the need to produce an estimate of the entire distribution of $p(x)$ at each value for $\theta$,
and the large ensemble sizes this implies,
and at this point it would be tempting to be put off.
However, we believe that there is a simple approximation that makes this potentially feasible.

\subsection{The assumption of normality}

The approximation that we propose to make the Jeffreys' Prior more tractable is to
assume normality (aka Gaussianity) for the distribution for $x$. We can then write $p(x|\theta)=p(x|\mu,\sigma^2,C)$, where
$\mu=\mu(\theta)$ and $\sigma^2=\sigma^2(\theta)$ are vectors of ensemble means and ensemble variances,
and $C=C(\theta)$ is a matrix of correlation coefficients.
The derivative terms in the definition of the Jeffreys' Prior then become derivatives of $\mu$, $\sigma$
and $C$, rather than derivatives in $\ln p$. These new derivatives can be evaluated
from numerical model integrations with much smaller ensembles than would be needed to
evaluate the derivatives in $\ln p$.

We believe this approximation is reasonable, since most climate models are validated against monthly, seasonal, annual or even decadal mean data, and such data is typically close to normally distributed.

\subsection{The assumption of independence}

Under the assumption of normality we believe it may be possible to write a closed-form expression for Jeffreys' Prior, although it
is somewhat difficult (and is a work in progress).
To simplify the problem, therefore, we also assume that the data used to validate the climate model are independent.
Whether this is true or not will vary from case to case, and depends on exactly what validation data is
 used and at what time intervals. But if true, Jeffreys' Prior can be
written very simply, and is derived below. To make the derivation easy to follow we derive four cases, in terms of increasing complexity, leading up to the most general case.

\subsubsection{Single observation, single parameter}

In this (artificially simple) case the probability $p(x|\theta)$ is given by:
\begin{eqnarray}
p(x|\theta)=\frac{1}{\sqrt{2\pi\sigma}} \mbox{exp}\left(-\frac{(x-\mu)^2}{2\sigma^2}\right)
\end{eqnarray}

where $x$ is the single observation we are validating against, $\theta$ is the single
parameter in the climate model, and $\mu(\theta)$ and $\sigma(\theta)$ are the ensemble mean
and ensemble standard deviation of initial condition ensembles as a function of $\theta$.

This gives:
\begin{eqnarray}
\ln p=-\ln \sqrt{2\pi} -\ln \sigma -\frac{(x-\mu)^2}{2\sigma^2}
\end{eqnarray}

Taking first derivatives wrt the parameter $\theta$ gives:
\begin{eqnarray}
\frac{\partial \ln p}{\partial \theta}
&=&-\frac{1}{\sigma}\frac{\partial \sigma}{\partial \theta}
   +\frac{(x-\mu)^2}{\sigma^3}\frac{\partial \sigma}{\partial \theta}
   +\frac{(x-\mu)}{\sigma^2}\frac{\partial \mu}{\partial \theta}
\end{eqnarray}

Taking second derivatives gives:
\begin{eqnarray}
\frac{\partial^2 \ln p}{\partial \theta^2}
&=&-\frac{1}{\sigma}\frac{\partial^2 \sigma}{\partial \theta^2}
   +\frac{1}{\sigma^2}\left(\frac{\partial \sigma}{\partial \theta}\right)^2\\\nonumber
 &&+\frac{(x-\mu)^2}{\sigma^3}\frac{\partial^2\sigma}{\partial \theta^2}
   -\frac{3(x-\mu)^2}{\sigma^4}\left(\frac{\partial \sigma}{\partial \theta}\right)^2
   -\frac{2(x-\mu)}{\sigma^3}\frac{\partial \sigma}{\partial \theta}\frac{\partial \mu}{\partial \theta}\\\nonumber
 &&+\frac{(x-\mu)}{\sigma^2}\frac{\partial^2\mu}{\partial \theta^2}
   -\frac{2(x-\mu)}{\sigma^3}\frac{\partial \mu}{\partial \theta}\frac{\partial \sigma}{\partial \theta}
   -\frac{1}{\sigma^2}\left(\frac{\partial \mu}{\partial \theta}\right)^2
\end{eqnarray}

Taking expectations over $x$, and using the defintions of $\mu$ and $\sigma$, which imply that:
\begin{eqnarray}
E(x-\mu)   &=& 0 \\
E(x-\mu)^2 &=& \sigma^2
\end{eqnarray}

this reduces this to:
\begin{eqnarray}
E\left(\frac{\partial^2 \ln p}{\partial \theta^2}\right)
&=&-\frac{2}{\sigma^2}\left(\frac{\partial \sigma}{\partial \theta}\right)^2
   -\frac{1}{\sigma^2}\left(\frac{\partial \mu}{\partial \theta}\right)^2
\end{eqnarray}

Jeffreys' Prior is thus given by:
\begin{eqnarray}
p(\theta)&=&\sqrt{-E\left(\frac{\partial^2 \ln p}{\partial \theta^2}\right)}\\
&=&\sqrt{\frac{2}{\sigma^2}\left(\frac{\partial \sigma}{\partial \theta}\right)^2
   +\frac{1}{\sigma^2}\left(\frac{\partial \mu}{\partial \theta}\right)^2}
\end{eqnarray}

One might additionally make the assumption that this expression is dominated by variations in the mean
rather than in the standard deviation, in which case this simplifies further to:

\begin{eqnarray}
p(\theta)=\frac{1}{\sigma}\left|\frac{\partial \mu}{\partial \theta}\right|
\end{eqnarray}

\subsubsection{Multiple observations, single parameter}

This case is very similar to the previous case, but slightly more realistic in that we now
have $n$ independent observations, rather than just 1:

\begin{eqnarray}
p(x|\theta)
&=&\prod_{i=1}^{n} \frac{1}{\sqrt{2\pi\sigma_i}} \mbox{exp}\left(-\frac{(x_i-\mu_i)^2}{2\sigma_i^2}\right)\\
\ln p
&=&\sum_{i=1}^{n} -\ln \sqrt{2\pi} -\ln \sigma_i -\frac{(x_i-\mu_i)^2}{2\sigma_i^2}\\
\frac{\partial \ln p}{\partial \theta}
&=&\sum_{i=1}^{n}-\frac{1}{\sigma_i}\frac{\partial \sigma_i}{\partial \theta}
   +\frac{(x_i-\mu_i)^2}{\sigma_i^3}\frac{\partial \sigma_i}{\partial \theta}
   +\frac{(x_i-\mu_i)}{\sigma_i^2}\frac{\partial \mu_i}{\partial \theta}\\
\frac{\partial^2 \ln p}{\partial \theta^2}
&=&\sum_{i=1}^{n}-\frac{1}{\sigma_i}\frac{\partial^2 \sigma_i}{\partial \theta^2}
   +\frac{1}{\sigma_i^2}\left(\frac{\partial \sigma_i}{\partial \theta}\right)^2\\\nonumber
 &&+\frac{(x_i-\mu_i)^2}{\sigma_i^3}\frac{\partial^2\sigma_i}{\partial \theta^2}
   -\frac{3(x_i-\mu_i)^2}{\sigma_i^4}\left(\frac{\partial \sigma_i}{\partial \theta}\right)^2
   -\frac{2(x_i-\mu_i)}{\sigma_i^3}\frac{\partial \sigma_i}{\partial \theta}\frac{\partial \mu_i}{\partial \theta}\\\nonumber
 &&+\frac{(x_i-\mu_i)}{\sigma_i^2}\frac{\partial^2\mu_i}{\partial \theta^2}
   -\frac{2(x_i-\mu_i)}{\sigma_i^3}\frac{\partial \mu_i}{\partial \theta}\frac{\partial \sigma_i}{\partial \theta}
   -\frac{1}{\sigma_i^2}\left(\frac{\partial \mu_i}{\partial \theta}\right)^2\\
E\left(\frac{\partial^2 \ln p}{\partial \theta^2}\right)
&=&\sum_{i=1}^{n}-\frac{2}{\sigma_i^2}\left(\frac{\partial \sigma_i}{\partial \theta}\right)^2
   -\frac{1}{\sigma_i^2}\left(\frac{\partial \mu_i}{\partial \theta}\right)^2\\
p(\theta)&=&\sqrt{\sum_{i=1}^{n}\frac{2}{\sigma_i^2}\left(\frac{\partial \sigma_i}{\partial \theta}\right)^2
   +\frac{1}{\sigma_i^2}\left(\frac{\partial \mu_i}{\partial \theta}\right)^2}
\end{eqnarray}

In the case in which $\sigma$ is assumed constant this becomes:
\begin{eqnarray}
p(\theta)&=&\sqrt{\sum_{i=1}^{n}\frac{1}{\sigma_i^2}\left(\frac{\partial \mu_i}{\partial \theta}\right)^2}
\end{eqnarray}

\subsubsection{Single observation, multiple parameters}

This case is very similar to the single observation single parameter case,
but now involves derivatives wrt pairs of parameters.
For clarity, we start by writing a pair of parameters as $(\theta,\phi)$:

\begin{eqnarray}
p(x|\theta,\phi)
&=&\frac{1}{\sqrt{2\pi\sigma}} \mbox{exp}\left(-\frac{(x-\mu)^2}{2\sigma^2}\right)\\
\ln p
&=&-\ln \sqrt{2\pi} -\ln \sigma -\frac{(x-\mu)^2}{2\sigma^2}\\
\frac{\partial \ln p}{\partial \theta}
&=&-\frac{1}{\sigma}\frac{\partial \sigma}{\partial \theta}
   +\frac{(x-\mu)^2}{\sigma^3}\frac{\partial \sigma}{\partial \theta}
   +\frac{(x-\mu)}{\sigma^2}\frac{\partial \mu}{\partial \theta}\\
\frac{\partial^2 \ln p}{\partial \theta \partial \phi}
&=&-\frac{1}{\sigma}\frac{\partial^2 \sigma}{\partial \theta \partial \phi}
   +\frac{1}{\sigma^2}\frac{\partial \sigma}{\partial \theta}\frac{\partial \sigma}{\partial \phi}\\\nonumber
 &&+\frac{(x-\mu)^2}{\sigma^3}\frac{\partial^2\sigma}{\partial \theta \partial \phi}
   -\frac{3(x-\mu)^2}{\sigma^4}\frac{\partial \sigma}{\partial \theta \partial \phi}
   -\frac{2(x-\mu)}{\sigma^3}\frac{\partial \sigma}{\partial \theta}\frac{\partial \mu}{\partial \phi}\\\nonumber
 &&+\frac{(x-\mu)}{\sigma^2}\frac{\partial^2\mu}{\partial \theta \partial \phi}
   -\frac{2(x-\mu)}{\sigma^3}\frac{\partial \mu}{\partial \theta}\frac{\partial \sigma}{\partial \phi}
   -\frac{1}{\sigma^2}\frac{\partial \mu}{\partial \theta}\frac{\partial \mu}{\partial \phi}\\
E\left(\frac{\partial^2 \ln p}{\partial \theta \partial \phi}\right)
&=&-\frac{2}{\sigma^2}\frac{\partial \sigma}{\partial \theta}\frac{\partial \sigma}{\partial \phi}
   -\frac{1}{\sigma^2}\frac{\partial \mu}{\partial \theta}\frac{\partial \mu}{\partial \phi}
\end{eqnarray}

If we now switch notation from using the two parameters $(\theta,\phi)$ to multiple
parameters $(\theta_1,...,\theta_m)$
Jeffreys' Prior is given by:

\begin{eqnarray}
p(\theta)&=&\sqrt{- \mbox{det} E\left(\frac{\partial^2 \ln p}{\partial \theta_i \partial \theta_j}\right) }\\
         &=&\sqrt{\mbox{det}\left(
         \frac{2}{\sigma^2}\frac{\partial \sigma}{\partial \theta_j}\frac{\partial \sigma}{\partial \theta_k}
         +\frac{1}{\sigma^2}\frac{\partial \mu}{\partial \theta_j}\frac{\partial \mu}{\partial \theta_k}\right)}
\end{eqnarray}

In the case where $\sigma$ is assumed constant this becomes:
\begin{eqnarray}
p(\theta)
         &=&\sqrt{\mbox{det}\left(
         \frac{\partial \mu}{\partial \theta_j}\frac{\partial \mu}{\partial \theta_k}\right)}\\
         &=&\frac{1}{\sigma}\sqrt{\mbox{det}\left(
         \frac{1}{\sigma^2}\frac{\partial \mu}{\partial \theta_j}\frac{\partial \mu}{\partial \theta_k}\right)}
\end{eqnarray}

\subsubsection{Multiple observations, multiple parameters}

This is the general case, and is the first case that could be applied to real climate models. Once again, we initially consider a pair of parameters $(\theta,\phi)$ initially.

\begin{eqnarray}
p(x|\theta,\phi)
&=&\prod_{i=1}^{n} \frac{1}{\sqrt{2\pi\sigma_i}} \mbox{exp}\left(-\frac{(x_i-\mu_i)^2}{2\sigma_i^2}\right)\\
\ln p
&=&\sum_{i=1}^{n}-\ln \sqrt{2\pi} -\ln \sigma_i -\frac{(x_i-\mu_i)^2}{2\sigma_i^2}\\
\frac{\partial \ln p}{\partial \theta}
&=&\sum_{i=1}^{n}-\frac{1}{\sigma_i}\frac{\partial \sigma_i}{\partial \theta}
   +\frac{(x_i-\mu_i)^2}{\sigma_i^3}\frac{\partial \sigma_i}{\partial \theta}
   +\frac{(x_i-\mu_i)}{\sigma_i^2}\frac{\partial \mu_i}{\partial \theta}\\
\frac{\partial^2 \ln p}{\partial \theta \partial \phi}
&=&\sum_{i=1}^{n}-\frac{1}{\sigma_i}\frac{\partial^2 \sigma_i}{\partial \theta \partial \phi}
   +\frac{1}{\sigma_i^2}\frac{\partial \sigma_i}{\partial \theta}\frac{\partial \sigma_i}{\partial \phi}\\\nonumber
 &&+\frac{(x_i-\mu_i)^2}{\sigma_i^3}\frac{\partial^2\sigma_i}{\partial \theta \partial \phi}
   -\frac{3(x_i-\mu_i)^2}{\sigma_i^4}\frac{\partial \sigma_i}{\partial \theta \partial \phi}
   -\frac{2(x_i-\mu_i)}{\sigma_i^3}\frac{\partial \sigma_i}{\partial \theta}\frac{\partial \mu_i}{\partial \phi}\\\nonumber
 &&+\frac{(x_i-\mu_i)}{\sigma_i^2}\frac{\partial^2\mu_i}{\partial \theta \partial \phi}
   -\frac{2(x_i-\mu_i)}{\sigma_i^3}\frac{\partial \mu_i}{\partial \theta}\frac{\partial \sigma_i}{\partial \phi}
   -\frac{1}{\sigma_i^2}\frac{\partial \mu_i}{\partial \theta}\frac{\partial \mu_i}{\partial \phi}\\
E\left(\frac{\partial^2 \ln p}{\partial \theta \partial \phi}\right)
&=&\sum_{i=1}^{n}-\frac{2}{\sigma_i^2}\frac{\partial \sigma_i}{\partial \theta}\frac{\partial \sigma_i}{\partial \phi}
   -\frac{1}{\sigma_i^2}\frac{\partial \mu_i}{\partial \theta}\frac{\partial \mu_i}{\partial \phi}
\end{eqnarray}

If we again switch notation from using the two parameters $(\theta,\phi)$ to multiple
parameters $(\theta_1,...,\theta_m)$
Jeffreys' Prior in this case is given by:

\begin{eqnarray}
p(\theta)&=&\sqrt{- \mbox{det} E\left(\frac{\partial^2 \ln p}{\partial \theta_j \partial \theta_k}\right) }\\
         &=&\sqrt{\mbox{det}\left(\sum_{i=1}^{n}
         \frac{2}{\sigma_i^2}\frac{\partial \sigma_i}{\partial \theta_j}\frac{\partial \sigma_i}{\partial \theta_k}
         +\frac{1}{\sigma_i^2}\frac{\partial \mu_i}{\partial \theta_j}\frac{\partial \mu_i}{\partial \theta_k}\right)}
\end{eqnarray}

In the case where $\sigma$ is assumed constant this becomes:
\begin{eqnarray}\label{eq35}
p(\theta)
         &=&\sqrt{\mbox{det}\left(\sum_{i=1}^{n}
         \frac{1}{\sigma_i^2}\frac{\partial \mu_i}{\partial \theta_j}\frac{\partial \mu_i}{\partial \theta_k}\right)}
\end{eqnarray}

\section{Summary and Discussion}\label{s5}

We have argued that Jeffreys' Prior may be a useful way to create probabilistic forecasts of future climate from numerical climate models, since
it is the least arbitrary choice of prior, and hence reduces the potential for argument about which prior is likely the best choice. It also allows for honest back-testing of climate models, unlike subjective priors.

Computing Jeffreys' priors for the general case of arbitrary distributions of modelled and observed variables is likely to be computationally too demanding, given
the computational cost of climate models. However, we have argued that by assuming that the variables against
which the model is validated are Gaussian the problem becomes tractable.
Using this approximation computing the Jeffreys' prior then involves computing derivatives of the mean, variance
and correlation coefficients of initial condition ensembles with respect to the parameters.
Under the further assumption that the observations are independent the correlation matrix becomes diagonal and
the computation of the prior reduces to evaluation of first derivatives of the mean and variance of initial condition
ensembles. This can be simplified even further by assuming that the variance is roughly constant, in which case the
prior becomes a function of the sensitivity of the ensemble mean to variations in the model parameters.

There are number of areas of further work, and a number of outstanding questions.

The most important next step is to attempt to apply Jeffreys' Prior, as approximate above, to
climate model results. We are trying this both using a simple energy balance model, and a fully complex
climate model. There are many practical questions related to implementation that we have not discussed here.

It would also be useful, if possible, to derive closed-form solutions for the case of correlated observations.

There are many outstanding questions. We are particularly interested in the question
of whether to use what we call a \emph{deterministic} or \emph{stochastic} approach to experimental design
for climate model integrations.
We distinguish between these approaches as follows.
Consider an experiment in which we use $n$ initial condition ensembles (each for fixed parameters), each
with $m$ members. In the limiting case as $m$ becomes very large, the initial condition uncertainty will
disappear entirely from the ensemble means and variances, and the ensemble means and variances become a
purely deterministic function of the parameters. Using large values of $m$ is therefore what we call the deterministic approach.
On the other hand, for $m=1$, the ensemble mean is affected by initial condition uncertainty, and is highly stochastic (i.e. not a deterministic function of the model parameters any more, although it does
contain a deterministic signal, obscured by the noise).
Using $m=1$ is what we call the stochastic (or stochastic parameter) approach.
We believe that there are statistical reasons why the
stochastic parameter approach is the most efficient, based on the
theory of experimental design, although the modelling of the variance response to changing parameters is
certainly then more complex.

Another set of related oustanding questions relates to emulators. It is now well accepted that emulators can,
in many cases, improve probabilistic predictions, including those from climate models. But can emulators
also help in evaluating the Jeffreys' Prior? Probably, but questions remain.
For instance: if an emulator is to be used to help evaluate equation~\ref{eq35}, then should the
emulator be applied before or after taking the derivative?

\appendix

\section{Proof that Jeffreys' Prior is the same under coordinate transformations, for a single parameter}

If there is just a single parameter $\theta$, then Jeffreys' Prior is defined as:
\begin{eqnarray}
    p(\theta)&=&\sqrt{-E\left[\frac{\partial^2 \ln p}{\partial \theta^2}\right]}\\
             &=&\sqrt{- \int \frac{\partial^2 \ln p}{\partial \theta^2} p(x,\theta) dx}
\end{eqnarray}

First, we will show that
\begin{equation}
    E\left[\frac{\partial^2 \ln p}{\partial \theta^2}\right]
    =
    -E\left[\left(\frac{\partial \ln p}{\partial \theta}\right)^2\right]
\end{equation}

To show this, we note that
\begin{eqnarray}
\frac{\partial \ln p}{\partial \theta}
&=&\frac{1}{p}\frac{\partial p}{\partial \theta}\\
\frac{\partial^2 \ln p}{\partial \theta^2}
&=&\frac{\partial}{\partial \theta} \left(\frac{1}{p}\frac{\partial p}{\partial \theta}\right)\\\nonumber
&=&\frac{1}{p}\frac{\partial^2 p}{\partial \theta^2}-\frac{1}{p^2}\left(\frac{\partial p}{\partial \theta}\right)^2\\\nonumber
&=&\frac{1}{p}\frac{\partial^2 p}{\partial \theta^2}-\left(\frac{\partial \ln p}{\partial \theta}\right)^2\\
\frac{p \partial^2 \ln p}{\partial \theta^2}
&=&\frac{\partial^2 p}{\partial \theta^2}-\left(p \frac{\partial \ln p}{\partial \theta}\right)^2
\end{eqnarray}

Integrating over all $x$, gives:
\begin{eqnarray}
E\left[\frac{\partial^2 \ln p}{\partial \theta^2}\right]
&=&-E\left[\left(\frac{\partial \ln p}{\partial \theta}\right)^2\right]
\end{eqnarray}

Given this result, Jeffreys' Prior can then be written as:
\begin{eqnarray}
    p(\theta)&=&\sqrt{E\left[\left(\frac{\partial \ln p}{\partial \theta}\right)^2\right]}\\
             &=&\sqrt{\int \left(\frac{\partial \ln p}{\partial \theta}\right)^2 p(x,\theta) dx}
\end{eqnarray}

Using this result a prediction based on Jeffreys' Prior
\begin{equation}
    p(y|x) \propto \int p(y|\theta) p(x|\theta) \sqrt{-E\left[\frac{\partial^2 \ln p}{\partial \theta^2}\right]} d \theta
\end{equation}

becomes
\begin{equation}
    p(y|x) \propto \int p(y|\theta) p(x|\theta) \sqrt{E\left[\left(\frac{\partial \ln p}{\partial \theta}\right)^2\right]} d \theta
\end{equation}

If we now change variables from $\theta$ to $\phi$, and apply standard rules for changing variables, we find,
for the first part of the integrand:

\begin{eqnarray}
p(y|\theta)&=&p(y|\phi)
\end{eqnarray}

For the second part of the integrand:
\begin{eqnarray}
p(x|\theta)&=&p(x|\phi)
\end{eqnarray}

For the third part of the integrand:

\begin{eqnarray}
\frac{\partial \ln p}{\partial \theta}&=&\frac{d \phi}{d \theta} \frac{\partial \ln p}{\partial \phi}\\
\left(\frac{\partial \ln p}{\partial \theta}\right)^2&=&\left(\frac{d \phi}{d \theta}\right)^2 \left(\frac{\partial \ln p}{\partial \phi}\right)^2\\
E\left[\left(\frac{\partial \ln p}{\partial \theta}\right)^2\right]
&=&\left(\frac{d \phi}{d \theta}\right)^2 E\left[\left(\frac{\partial \ln p}{\partial \phi}\right)^2\right]\\
\left(E\left[\left(\frac{\partial \ln p}{\partial \theta}\right)^2\right]\right)^{\frac{1}{2}}
&=&\left(\frac{d \phi}{d \theta}\right) \left(E\left[\left(\frac{\partial \ln p}{\partial \phi}\right)^2\right]\right)^{\frac{1}{2}}
\end{eqnarray}

and for the fourth part of the integrand:
\begin{eqnarray}
d\theta&=&\frac{d \theta}{d \phi} d\phi
\end{eqnarray}

Putting this all together, the prediction becomes:
\begin{eqnarray}
p(y|x)
&\propto& \int p(y|\theta) p(x|\theta) \left(E\left[\left(\frac{\partial \ln p}{\partial \theta}\right)^2\right]\right)^{\frac{1}{2}} d\theta\\
&\propto& \int p(y|\phi) p(x|\phi) \left(\frac{d \phi}{d \theta}\right) \left(E\left[\left(\frac{\partial \ln p}{\partial \phi}\right)^2\right]\right)^{\frac{1}{2}} \frac{d \theta}{d \phi} d\phi\\
&\propto& \int p(y|\phi) p(x|\phi) \left(E\left[\left(\frac{\partial \ln p}{\partial \phi}\right)^2\right]\right)^{\frac{1}{2}} d\phi
\end{eqnarray}

and so:
\begin{equation}
    p(y|x) \propto \int p(y|\phi) p(x|\phi) \sqrt{E\left[\left(\frac{\partial \ln p}{\partial \phi}\right)^2\right]} d \phi
\end{equation}

We see that we would have achieved the same result had we parametrised using $\phi$ in the first place.
In other words it does not matter which coordinates we choose: the prediction will always be the same.

\section{Proof that Jeffreys' Prior is the same under coordinate transformations, for multiple parameters}

For multiple parameters $\theta$ Jeffreys' Prior is defined as:
\begin{eqnarray}
    p(\theta)&=&\sqrt{-\mbox{det} \left(E\left[\frac{\partial^2 \ln p}{\partial \theta_i \theta_j}\right]\right)}
\end{eqnarray}

First, we will show that, for a pair of parameters $(\theta,\phi)$:
\begin{eqnarray}
E\left[\frac{\partial^2 \ln p}{\partial \theta \partial \phi}\right]
&=&-E\left[\frac{\partial \ln p}{\partial \theta}\frac{\partial \ln p}{\partial \phi}\right]
\end{eqnarray}

To show this, we note that
\begin{eqnarray}
\frac{\partial \ln p}{\partial \theta}
&=&\frac{1}{p}\frac{\partial p}{\partial \theta}\\
\frac{\partial^2 \ln p}{\partial \theta \partial \phi}
&=&\frac{\partial}{\partial \phi} \left(\frac{1}{p}\frac{\partial p}{\partial \theta}\right)\\\nonumber
&=&\frac{1}{p}\frac{\partial^2 p}{\partial \theta \partial \phi}-\frac{1}{p^2}\frac{\partial p}{\partial \theta}\frac{\partial p}{\partial \phi}\\\nonumber
&=&\frac{1}{p}\frac{\partial^2 p}{\partial \theta \partial \phi}-\frac{1}{p^2}\frac{\partial \ln p}{\partial \theta}\frac{\partial \ln p}{\partial \phi}\\
\frac{p \partial^2 \ln p}{\partial \theta \partial \phi}
&=&\frac{\partial^2 p}{\partial \theta \partial \phi}-p \frac{\partial \ln p}{\partial \theta}\frac{\partial \ln p}{\partial \phi}
\end{eqnarray}

Integrating over all $x$, gives:

\begin{eqnarray}
E\left[\frac{\partial^2 \ln p}{\partial \theta \partial \phi}\right]
&=&-E\left[\frac{\partial \ln p}{\partial \theta}\frac{\partial \ln p}{\partial \phi}\right]
\end{eqnarray}

Given this result, Jeffreys' Prior can then be written as:
\begin{eqnarray}
    p(\theta)&=&\sqrt{\mbox{det} \left(E\left[\frac{\partial \ln p}{\partial \theta}\frac{\partial \ln p}{\partial \phi}\right]\right)}
\end{eqnarray}

Using this result a prediction based on Jeffreys' Prior
\begin{equation}
    p(y|x) \propto \int p(y|\theta) p(x|\theta) \sqrt{-\mbox{det} E\left[\frac{\partial^2 \ln p}{\partial \theta_i \partial \theta_j }\right]} d \theta_1 ... d \theta_n
\end{equation}

becomes
\begin{equation}
    p(y|x) \propto \int p(y|\theta) p(x|\theta) \sqrt{\mbox{det}
    E\left[\frac{\partial \ln p}{\partial \theta_i}\frac{\partial \ln p}{\partial \theta_j}\right]} d \theta_1 ... d \theta_n
\end{equation}

If we now change variables from $\theta$ to $\phi$, and apply standard rules for changing variables, we find,
for the first part of the integrand:

\begin{eqnarray}
p(y|\theta)&=&p(y|\phi)
\end{eqnarray}

For the second part of the integrand:
\begin{eqnarray}
p(x|\theta)&=&p(x|\phi)
\end{eqnarray}

For the third part of the integrand:

\begin{eqnarray}
\frac{\partial \ln p}{\partial \theta_i}&=&\frac{d \phi_k}{d \theta_i} \frac{\partial \ln p}{\partial \phi_k}\\
\frac{\partial \ln p}{\partial \theta_j}&=&\frac{d \phi_l}{d \theta_j} \frac{\partial \ln p}{\partial \phi_l}\\
\frac{\partial \ln p}{\partial \theta_i}\frac{\partial \ln p}{\partial \theta_j}
&=&\frac{d \phi_k}{d \theta_i} \frac{\partial \ln p}{\partial \phi_k}\frac{d \phi_l}{d \theta_j} \frac{\partial \ln p}{\partial \phi_l}\\
&=&\frac{d \phi_k}{d \theta_i}\frac{d \phi_l}{d \theta_j}\frac{\partial \ln p}{\partial \phi_k}\frac{\partial \ln p}{\partial \phi_l}\\
E\left[\frac{\partial \ln p}{\partial \theta_i}\frac{\partial \ln p}{\partial \theta_j}\right]
&=&\frac{d \phi_k}{d \theta_i}\frac{d \phi_l}{d \theta_j} E\left[\frac{\partial \ln p}{\partial \phi_k}\frac{\partial \ln p}{\partial \phi_l}\right]\\
\mbox{det} E\left[\frac{\partial \ln p}{\partial \theta_i}\frac{\partial \ln p}{\partial \theta_j}\right]
&=&\mbox{det} \left(\frac{d \phi_k}{d \theta_i}\frac{d \phi_l}{d \theta_j} E\left[\frac{\partial \ln p}{\partial \phi_k}\frac{\partial \ln p}{\partial \phi_l}\right]\right)\\
&=&\mbox{det} \left(\frac{d \phi_k}{d \theta_i}\frac{d \phi_l}{d \theta_j} \right)
  \mbox{det} \left(E\left[\frac{\partial \ln p}{\partial \phi_k}\frac{\partial \ln p}{\partial \phi_l}\right]\right)\\
&=&\mbox{det} \left(\frac{d \phi_k}{d \theta_i}\right) \mbox{det}\left(\frac{d \phi_l}{d \theta_j} \right)
  \mbox{det} \left(E\left[\frac{\partial \ln p}{\partial \phi_k}\frac{\partial \ln p}{\partial \phi_l}\right]\right)\\
&=&J(\phi,\theta) J(\phi,\theta)
  \mbox{det} \left(E\left[\frac{\partial \ln p}{\partial \phi_k}\frac{\partial \ln p}{\partial \phi_l}\right]\right)
\end{eqnarray}

(where $J(\theta,\phi)$ is the Jacobian determinant)

and for the fourth part of the integrand:
\begin{eqnarray}
d\theta_1...d\theta_n&=&J(\theta,\phi) d\phi_1...d\phi_n
\end{eqnarray}

Putting this all together, the prediction becomes:
\begin{eqnarray}
p(y|x)
&\propto& \int p(y|\theta) p(x|\theta) \left(\mbox{det} E\left[\frac{\partial \ln p}{\partial \theta_i}\frac{\partial \ln p}{\partial \theta_j}\right]\right)^{\frac{1}{2}} d\theta_1...d\theta_n\\
&\propto& \int p(y|\phi) p(x|\phi) J(\phi,\theta) \left(E\left[\frac{\partial \ln p}{\partial \phi_k}\frac{\partial \ln p}{\partial \phi_l}\right]\right)^{\frac{1}{2}} J(\theta,\phi) d\phi_1...d\phi_n\\
&\propto& \int p(y|\phi) p(x|\phi) \left(E\left[\frac{\partial \ln p}{\partial \phi_k}\frac{\partial \ln p}{\partial \phi_l}\right]\right)^{\frac{1}{2}} d\phi_1...d\phi_n\\
\end{eqnarray}

We see that also in this case we would have achieved the same result had we parametrised using $\phi$ in the first place.

\bibliography{arxiv}


\end{document}